# Nontrivial Berry Phase and Type-II Dirac Transport in Layered Material PdTe$_2$

Fucong Fei[1], Xiangyan Bo[1], Rui Wang[1], Bin Wu[1], Juan Jiang[2], Dongzhi Fu[1], Ming Gao[3], Hao Zheng[4], Yulin Chen[2,5], Xuefeng Wang[3], Haijun Bu[1], Fengqi Song[1,*], Xiangang Wan[1,†], Baigeng Wang[1,‡], Guanghou Wang[1]

[1] *National Laboratory of Solid State Microstructures, Collaborative Innovation Center of Advanced Microstructures, and College of Physics, Nanjing University, Nanjing, 210093, P. R. China*

[2] *School of Physical Science and Technology, ShanghaiTech University, and CAS-Shanghai Science Research Center, Shanghai 200031, P. R. China*

[3] *National Laboratory of Solid State Microstructures, Collaborative Innovation Center of Advanced Microstructures, and School of Electronic Science and Engineering, Nanjing University, Nanjing, 210093, P. R. China*

[4] *Department of Physics and Astronomy, Shanghai Jiao Tong University, Shanghai 200240, P. R. China*

[5] *Department of Physics, University of Oxford, Oxford, OX1 3PU, UK*

# ABSTRACT

Here we report the evidence of the type-II Dirac Fermion in the layered crystal PdTe$_2$. The de Haas-van Alphen oscillations find a small Fermi pocket with a cross section of 0.077nm$^{-2}$ with a nontrivial Berry phase. First-principal calculations reveal that it is originated from the hole pocket of a tilted Dirac cone. Angle Resolved Photoemission Spectroscopy demonstrates a type-II Dirac cone featured dispersion. We also suggest PdTe$_2$ is an improved platform to host the topological superconductors.

In condensed matters, multiple excitations can be implemented to simulate the physics of new particles, such as Dirac fermions [1-10], Weyl fermions [11-24], Majorana fermions [25-28] and exotic new fermions beyond Dirac and Weyl fermions [29] with the potential applications in next-generation spintronics and quantum computing. These particles might be unrealistic in real world while allowable in condensed matters due to versatile operations of symmetry breaking. In Dirac semimetals, e. g., $Na_3Bi$ [8-10] and $Cd_3As_2$ [1-7], topological protected gapless Dirac cones with liner dispersions are discovered and electrons in these materials can be described by massless Dirac equation. Interesting negative magnetoresistance appears, which simulates the chiral anomaly in high energy physics. By breaking inversion symmetry or time reversal symmetry, a Dirac cone can split into a pair of Weyl cones and form Weyl semimetals such as TaAs family [12-17]. Noncontinuous Fermi arcs appear on their surfaces. In several Weyl semimetals, for instance $Mo_xW_{1-x}Te_2$ [19-23] and LaAlGe [11], the Lorentz invariance is broken and the Weyl cones are strongly tilted to form type-II Weyl semimetals, which are predicted to have unique properties such as field-selective anomaly and chiral mode reversal [30], as well as Klein tunneling in momentum space [31]. Recently, the Lorentz invariance breaking is suggested in Dirac semimetals with the result of type-II Dirac fermions with tilted Dirac cone in $PtSe_2$ family [32, 33].

Here we demonstrate a nontrivial Berry phase in a layered material $PdTe_2$ by using the de Haas-van Alphen (dHvA) oscillations for the first time. Six conductive pockets are identified in the dHvA measurements, where the α mode with the frequency of 8.0T exhibits the nontrivial Berry phase. The calculations confirm the Berry phase is originated from the hole pocket of a tilted type-II Dirac cone. This is also confirmed by the dispersions in Angle Resolved Photoemission Spectroscopy (ARPES). Remembering its anisotropic superconductivity under 1.9K, we suggest $PdTe_2$ might be an improved platform to host the topological superconductors (TSCs).

Single crystal $PdTe_2$ was grown by the melt-growth method. A certain amount of Pd and Te powder (from Alfa Aesar) with atomic ratio 1:2.2 was sealed in an evacuated ampoule. The ampoule was heated up and maintained at 790℃ for 48 hours. The melt was then slowly cooled down to 500℃ in 7 days and then annealed at this temperature for 7 more days before natural cooling to room temperature. Millimeter-sized crystals with metallic luster were obtained, which could be exfoliated to small flakes by a knife easily. As shown in Fig. 1(a), $PdTe_2$ is a type of layered transition metal dichalcogenides (TMD) material of $CdI_2$-type structure with the $P\bar{3}m1$ space group. The x-ray diffraction data of the single crystal is shown in Fig. 1(b). The strong (00n) peaks of $PdTe_2$ can be clearly seen and no other impurity peaks can be found, indicating nice crystallization of the $PdTe_2$ sample. Fig. 1(c) shows the energy dispersive spectra for a typical $PdTe_2$ flake. The ratio between Pd and Te elements is 1:1.99, indicating the stoichiometric ratio of the sample. The EDS mapping (insets of Fig. 1(c)) show the uniform elemental distribution of Pd and Te, respectively.

We measure the magnetoresistance (MR) curves of $PdTe_2$ single crystal under low temperatures. The longitudinal resistance under 2K is shown in Fig. 1(d), where no regular oscillations can be identified. Because of the multiple complex conductive pockets in PdTe$_2$ [34], it is hard to identify which bands dominate the electric transport. Different bands may contribute different transport parameters such as carrier density, mobility and Landau fill factor. The oscillations in electrical transport are reliable on these parameters. Hence the absence of Shubnikov-de Haas oscillations in electrical transport is reasonable, which also fail in other reports [35]. The inset in Fig. 1(d) shows the hall resistance under various temperatures from 2 K to 150 K. One can clearly see the hall resistance versus magnetic field is linear at 150K but becomes bended when the temperature goes down to 2K, indicating the multiple band transport with complex contribution from various conductive pockets near the Fermi surface. The sudden drop of both R$_{xx}$ and R$_{xy}$ near zero fields at 2K is caused by the superconducting transition as described below.

PdTe$_2$ is a superconductor with $T_c$ ~ 2 K [36]. We make further study on the superconductivity property of PdTe$_2$. The crystal shows an anisotropic superconductivity as seen when we measure the temperature dependent resistance under magnetic field using two different sets of devices of configuration 1 (the current is perpendicular to c axis; the field is parallel to c axis) and 2 (the current is parallel to c axis; the field is parallel to c axis), as seen in Fig. 1(e) and (f) respectively. Using configuration 2, with increasing the field to 500 Oe, the superconductivity is nearly suppressed while the superconductivity survives the field of 2000 Oe in configuration 1. We can also see that in Fig. 1(f), the resistance increases when temperature goes down near $T_c$, which is absent in Fig. 1(e). The anisotropic superconductivity in PdTe$_2$ is thus obvious.

The nontrivial Berry phase is demonstrated by the dHvA measurements after we measure the magnetization condition of the PdTe$_2$ flakes under low temperatures. Fig. 2(a) displays the magnetization versus magnetic field (B // c) for PdTe$_2$ at 1.8K. Unexpectedly and Contrary to the MR measurements, beautiful dHvA oscillations can be seen from the raw data of M-H curves. Fig. 2(b) shows the magnetization strength versus 1/B under various temperatures after the background subtraction. Multiple sets of oscillations, with the frequencies of 8.0 (α), 113.2 (β), 117.9 (γ), 124.3 (δ), 133.9 (ε) and 455.8T (ζ), can be extracted from the M-H data by the fast Fourier transformation (FFT) (inset in Fig. 2(c)). The dHvA oscillation can be described by the Lifshitz-Kosevich (LK) formula:

$$\Delta M \propto -R_T R_D \sin[2\pi(\frac{F}{B}-(\frac{1}{2}-\phi))] \qquad (1)$$

The thermal damping factor is $R_T = \frac{\chi T}{\sinh(\chi T)}$ and the Dingle damping factor is $R_D = \exp(-\chi \cdot T_D)$ where $\chi = \frac{2\pi^2 k_B m^*}{\hbar eB}$. $\Phi$ is the phase shift and $\phi = \frac{\phi_B}{2\pi} - \delta$, where $\Phi_B$ is the Berry phase and $\delta$ equals 0 and $\pm 1/8$ for two dimensional and three dimensional systems respectively. The effective mass $m^*$ can be extracted by the fit of the temperature dependence of the corresponding oscillation amplitude to the $R_T$. The Dingle temperature $T_D$ can also be obtained through the fit of Dingle damping factor $R_D$, meanwhile, quantum relaxation time $\tau$ can be calculated by $\tau = \frac{\hbar}{2\pi k_B T_D}$ and quantum mobility $\mu_q = e\tau/m^*$. The Berry phases determine the topological properties of the conductive pockets and can be extracted using Landau level fan diagram. In dHvA oscillation, the integer landau index should be assigned when density of states (DOS) of Fermi surface, which is proportion to $dM/dB$, reaches a minimum. Therefore, the landau index of the dHvA oscillation minima should be n-1/4 [37]. We find α mode can be separated easily, while the four mid-frequency modes (β, γ, δ, ε) are with similar frequencies. In addition, oscillation of high-frequency mode (ζ) is masked by other oscillations because of the small amplitude. Therefore, we fit the oscillatory components of these five modes by the multiband LK formula, as shown in Fig. 2(e), and extract the transport parameters. All six sets of data of the dHvA oscillations are displayed in Table I.

Interestingly, a topological-nontrivial mode, the α mode, is evidenced by the Landau fan diagram, by using dHvA oscillation minima as the landau index of n-1/4, as shown in Fig. 2(d). The intercept of the linear fitting is 0.46, which is the signature of nontrivial Dirac transport. Fig. 2(c) displays the fits of effective mass of the all six oscillatory modes by the corresponding FFT amplitudes versus temperature (inset in Fig. 2(c)). It is clear that the effective mass of the low-frequency mode (α) is much less than other modes. The low effective mass also agrees with the Dirac nature of the α mode.

To identify the contributing electronic pocket of each oscillation mode, the assistance from the band structural calculation is needed. We calculate the band structure of $PdTe_2$ (Fig. 3(a)) and the Brillouin zone (BZ) is shown in Fig. 3(b), where high-symmetry points, lines and Dirac point (D) are also indicated. There is a band crossing feature near the Fermi level along the Γ-A line. This band crossing is unavoidable, because these two bands belong to different representations (G4 and G5+G6 respectively). This is determined by the C3 rotational symmetry around the c axis [32, 33, 38]. Remarkably, these two bands show linear dispersions in the vicinity of the Dirac point along both the in-plane ($k_x$-$k_y$) and out-of-plane ($k_z$) directions and the Dirac cone is untilted along $k_x$-$k_y$ plane (Fig. 3(e)) but tilted strongly along ($k_z$) direction (Fig. 3(f)), which is the characteristic feature of the type-II Dirac fermions as reported [32, 33]. By our calculations, the Dirac point is at k = (0,0,±0.40).

In the calculation, a series of electronic pockets can be seen on the Fermi surface as shown in Fig. 3(c) and (d). For figuring out the contributing pocket of each dHvA oscillation mode, we calculate the extremal surfaces of each pocket along kz direction. The modes around 100T were considered contributing from the minimum extremal surfaces (yellow area in Fig. 3(d)) of six pliers shaped pockets (translucent blue pockets in Fig. 3(d)) at kz=±0.5. Departure from high symmetry points and complex contours of these pockets may explain the several beat frequencies around 100T. The extremal surface of purple colored pocket at K point in Fig. 3(d) is about 0.04 Å$^{-2}$, which is the possible origin of ζ mode (456T).

In type-II Dirac semimetals, the Dirac cone is tilted strongly because of the Lorentz invariance breaking, which causes the formation of a pair of electron and hole pockets[39]. In PdTe$_2$, the Fermi level is about 0.53 eV above the Dirac point. Thus the hole pocket shrinks to a small pocket at Γ point, as seen in Fig. 3(a) (also marked by red dash line in Figs. 3(c, d)) while the electron pocket, which corresponding to the translucent-aqua-like pocket in Figs. 3(c, d), becomes much bigger and forms an apple-pit-shaped pocket in reciprocal space. We find that the cut of the small hole pocket is with the similar area to that of the nontrivial α mode (8T), which further confirms the Dirac nature of this pocket. For the big electron pocket, though the shape is complex, we verify that only a single maximum cross section achieves when $k_z=0$ after precise analysis. The corresponding oscillation frequency contribution is over 5kT, which is too high to detect in our experiment.

The type-II Dirac dispersion in $PdTe_2$ is characterized by the ARPES measurement. Fig. 3(g) shows the electronic dispersion along T-D direction. The linear dispersion of Dirac cone can be clearly seen. Noting that T-D line is perpendicular to $k_z$, the Dirac cone is not tilted as predicted. The evolution of the Dirac cone dispersion can be clearly seen under different photon energy (Figure S1) which reveals the bulk properties of the Dirac cone. Though it is difficult to detect dispersion along $k_z$ direction in $PdTe_2$, such a kind of layered material, we use different photon energy to collect dispersion at $k_y=0$ and map out the $k_x$-$k_z$ constant energy contours (Figure S2). A hole pocket and an electron pocket can be seen and tend to touch each other at Dirac point when energy goes down, which is a character of type-II Dirac dispersion.

We suggest the discovered type-II Dirac semimetal be an improved platform of TSC, on which a lot of efforts are made [27, 40-44]. Recently, with the discovery of three-dimensional topological semimetal, the attempts are made on inducing the SWS in the newly discovered semimetals [45, 46]. We here compare the TSC transport based on three kinds of mother materials, Weyl semimetal, Dirac semimetal and type-II Dirac semimetal. For the TSC arisen from the Weyl semimetals, it is known from the theoretical work [45, 46] that the chiral anomaly survived and invokes a Fermi arc surface state, in which the electrons remain unpaired and no gap opens. Meanwhile, the effective $p_x+ip_y$ pairing state in the bulk leads to chiral Majorana surface mode. The coexistence of Fermi arc and the Majorana mode may hamper its application in quantum computation. For the TS arisen from Dirac semimetals, the gapless node is Dirac point which is fourfold degenerated and additional space group symmetries are required for the stability of the Dirac nodes. The Fermi arc state of the Dirac semimetal is not topologically stable [47]. Hence, after the superconductivity sets in, in general, the surface electrons would form Cooper pairs and gap out the surface state. The situation is improved for the case of type-II Dirac cone where the dispersion is tilted as shown in Fig. 4. The significant superiority can be seen is that the electron and hole pockets near the Dirac points provide the plenty density of states which is favorable for both superconducting and TSC carrier ratio. This also means larger carrier density in the TS arisen from type-II Dirac semimetals.

$PdTe_2$ verified by our work to be a type-II Dirac semimetal with nontrivial Berry phase and an anisotropic superconductor below 1.9K. It is therefore a possible improved platform to search for mysterious Majorana Fermions and applies to the next-generation spintronics devices.

ACKNOWLEDGEMENTS

We gratefully acknowledge the financial support of the National Key Projects for Basic Research of China (Grant Nos: 2013CB922103), the National Natural Science Foundation of China (Grant Nos: 91622115, 91421109, 11574133 and 11274003), the PAPD project, the Natural Science Foundation of Jiangsu Province (Grant BK20130054), and the Fundamental Research Funds for the Central Universities. Use of the Stanford Synchrotron Radiation Lightsource, SLAC National Accelerator Laboratory, is supported by the U.S. Department of Energy, Office of Science, Office of Basic Energy Sciences under Contract No. DE-AC02-76SF00515. The technical support from the Hefei National Synchrotron Radiation Laboratory is acknowledged. We also thank Prof. Yongchun Tao in Nanjing Normal University, China for stimulating discussions.

## FOOTNOTES AND REFERENCE CITATION

The first three authors (F. F, X. B and R. W) contributed equally to this work.

[*] Corresponding authors.

songfengqi@nju.edu.cn

[†] Corresponding authors.

xgwan@nju.edu.cn

[‡] Corresponding authors.

bgwang@nju.edu.cn

[1] Z. K. Liu *et al.*, Nat Mater **13**, 677 (2014).
[2] L. P. He, X. C. Hong, J. K. Dong, J. Pan, Z. Zhang, J. Zhang, and S. Y. Li, Phys. Rev. Lett. **113**, 246402 (2014).
[3] H. Wang *et al.*, Nat. Mater. **15**, 38 (2016).
[4] M. Neupane *et al.*, Nat. Commun. **5**, 3786 (2014).
[5] S. Jeon *et al.*, Nat. Mater. **13**, 851 (2014).
[6] L. He, Y. Jia, S. Zhang, X. Hong, C. Jin, and S. Li, npj Quantum Mater. **1**, 16014 (2016).
[7] S. Borisenko, Q. Gibson, D. Evtushinsky, V. Zabolotnyy, B. Buchner, and R. J. Cava, Phys. Rev. Lett. **113**, 027603 (2014).
[8] S.-Y. Xu *et al.*, Science **347**, 294 (2015).
[9] J. Xiong, S. K. Kushwaha, T. Liang, J. W. Krizan, M. Hirschberger, W. Wang, R. J. Cava, and N. P. Ong, Science **350**, 413 (2015).
[10] Z. K. Liu *et al.*, Science **343**, 864 (2014).
[11] S.-Y. Xu *et al.*, arXiv, 1603.07318 (2016).
[12] L. X. Yang *et al.*, Nat. Phys. **11**, 728 (2015).
[13] H. M. Weng, C. Fang, Z. Fang, B. A. Bernevig, and X. Dai, Phys. Rev. X **5**, 011029 (2015).
[14] N. Xu *et al.*, Nat. Commun. **7**, 11006 (2016).
[15] B. Q. Lv *et al.*, Phys. Rev. X **5**, 031013 (2015).
[16] S.-Y. Xu *et al.*, Science **349**, 613 (2015).
[17] H. Zheng *et al.*, ACS Nano **10**, 1378 (2016).
[18] Y. Qi *et al.*, Nat. Commun. **7**, 11038 (2016).
[19] L. Huang *et al.*, Nat. Mater. **15**, 1155 (2016).
[20] X. C. Pan *et al.*, Nat. Commun. **6**, 7805 (2015).
[21] Z. J. Wang, D. Gresch, A. A. Soluyanov, W. W. Xie, S. Kushwaha, X. Dai, M. Troyer, R. J. Cava, and B. A. Bernevig, Phys. Rev. Lett. **117**, 056805 (2016).
[22] I. Belopolski *et al.*, Phys. Rev. B **94**, 085127 (2016).
[23] I. Belopolski *et al.*, Nat. Commun. **7**, 13643 (2016).
[24] Y. Xu, F. Zhang, and C. Zhang, Phys. Rev. Lett. **115**, 265304 (2015).
[25] L. Fu and C. L. Kane, Phys. Rev. Lett. **100**, 096407 (2008).
[26] J. J. He, T. K. Ng, P. A. Lee, and K. T. Law, Phys. Rev. Lett. **112**, 037001 (2014).
[27] H.-H. Sun *et al.*, Phys. Rev. Lett. **116**, 257003 (2016).
[28] J. P. Xu *et al.*, Phys. Rev. Lett. **114**, 017001 (2015).
[29] B. Bradlyn, J. Cano, Z. Wang, M. G. Vergniory, C. Felser, R. J. Cava, and B. A. Bernevig, Science **353**, 6299 (2016).
[30] M. Udagawa and E. J. Bergholtz, Phys. Rev. Lett. **117**, 086401 (2016).
[31] T. E. O'Brien, M. Diez, and C. W. Beenakker, Phys. Rev. Lett. **116**, 236401 (2016).

[32] H. Huang, S. Zhou, and W. Duan, Phys. Rev. B **94**, 121117(R) (2016).
[33] M. Yan *et al.*, arXiv, 1607.03643 (2016).
[34] A. E. Dunsworth, J. Low Temp. Phys. **19**, 51 (1975).
[35] Y. Wang *et al.*, Sci. Rep. **6**, 31554 (2016).
[36] v. J. Guggenheim and F. H. u. J. Müller, H. P. A. **34**, 408 (1961).
[37] J. Hu *et al.*, Phys. Rev. Lett. **117**, 016602 (2016).
[38] Y. Liu *et al.*, Chinese Phys. Lett. **32**, 067303 (2015).
[39] A. A. Soluyanov, D. Gresch, Z. Wang, Q. Wu, M. Troyer, X. Dai, and B. A. Bernevig, Nature **527**, 495 (2015).
[40] S.-Y. Xu *et al.*, Nat. Phys. **10**, 943 (2014).
[41] M.-X. Wang *et al.*, Science **336**, 52 (2012).
[42] S. Yonezawa, K. Tajiri, S. Nakata, Y. Nagai, Z. Wang, K. Segawa, Y. Ando, and Y. Maeno, Nat. Phys., doi:10.1038/nphys3907 (2016).
[43] Y. S. Hor *et al.*, Phys. Rev. Lett. **104**, 057001 (2010).
[44] Z. Liu, X. Yao, J. Shao, M. Zuo, L. Pi, S. Tan, C. Zhang, and Y. Zhang, Journal of the American Chemical Society **137**, 10512 (2015).
[45] G. Bednik, A. A. Zyuzin, and A. A. Burkov, Phys. Rev. B **92**, 035153 (2015).
[46] R. Wang, L. Hao, B. Wang, and C. S. Ting, Phys. Rev. B **93**, 184511 (2016).
[47] M. Kargarian, M. Randeria, and Y. M. Lu, arXiv, 1509.02180 (2016).

FIGURE CAPTIONS

FIG. 1. PdTe$_2$ Crystal growth and electrical transport under magnetic fields. (a) The layered structure (CdI2-type) of PdTe$_2$ crystals. (b) The X-ray diffraction data. The inset shows the optical micrograph of several flakes. (c) The EDS spectrum. The insets show the elemental distribution of Pd and Te, respectively, of a typical flake obtained by the EDS mapping. (d) The magnetoresistance of PdTe$_2$ sample under 2K. The inset is Hall resistance versus magnetic field under various temperatures. (e) and (f) are respectively temperature dependence of resistance under different fields of configuration 1 (I⊥c, B∥c) and 2 (I∥c, B∥c).

FIG. 2. The dHvA oscillations and nontrivial Berry phase. (a) The magnetization curve (B∥c) for PdTe$_2$ at 1.8K. (b) The magnetization strength versus 1/B under various temperatures after background subtraction. (c) The fits of effective mass for all six oscillation modes (α to ζ). The inset is the FFT amplitudes versus temperature. Peak corresponding to each oscillation mode was marked by Greek letter. (d) The Landau fan diagram for the dHvA oscillation of low frequency mode α. The inset is the fit of Dingle temperature ($T_D$). (e) Oscillatory component of other frequencies (β to ζ) and the multiband LK fit of it.

FIG. 3. Matching the dHvA components in the calculations of type-II Dirac cone. (a) The calculated electronic structures plotted along the directions shown in panel b. (b) The Brillouin zone and the high symmetry axis. (c) and (d) show the contour of Fermi

level along kz and kx-ky plane respectively. Pockets irrelevant to the hole and electron pockets form by the tilted Dirac cone in (d) are concealed for clearness. (e) and (f) show the projection of the Dirac cone along xy and xz direction respectively. (g) The dispersion along T-D direction measured by ARPES.

FIG. 4. Optimizing the topological superconductor carrier (TSC) by three-dimensional semimetals. This figure displays the band structure diagrams of the possible topological superconductors based on the mother materials of type-II Dirac semimetal when $T<T_c$. The red solid cross curves in superconducting (SC) gap (grey zone) represent the Majorana chiral mode. The Fermi arc states are killed in Dirac semimetals when the superconductivity occurs.

**FIGURES**

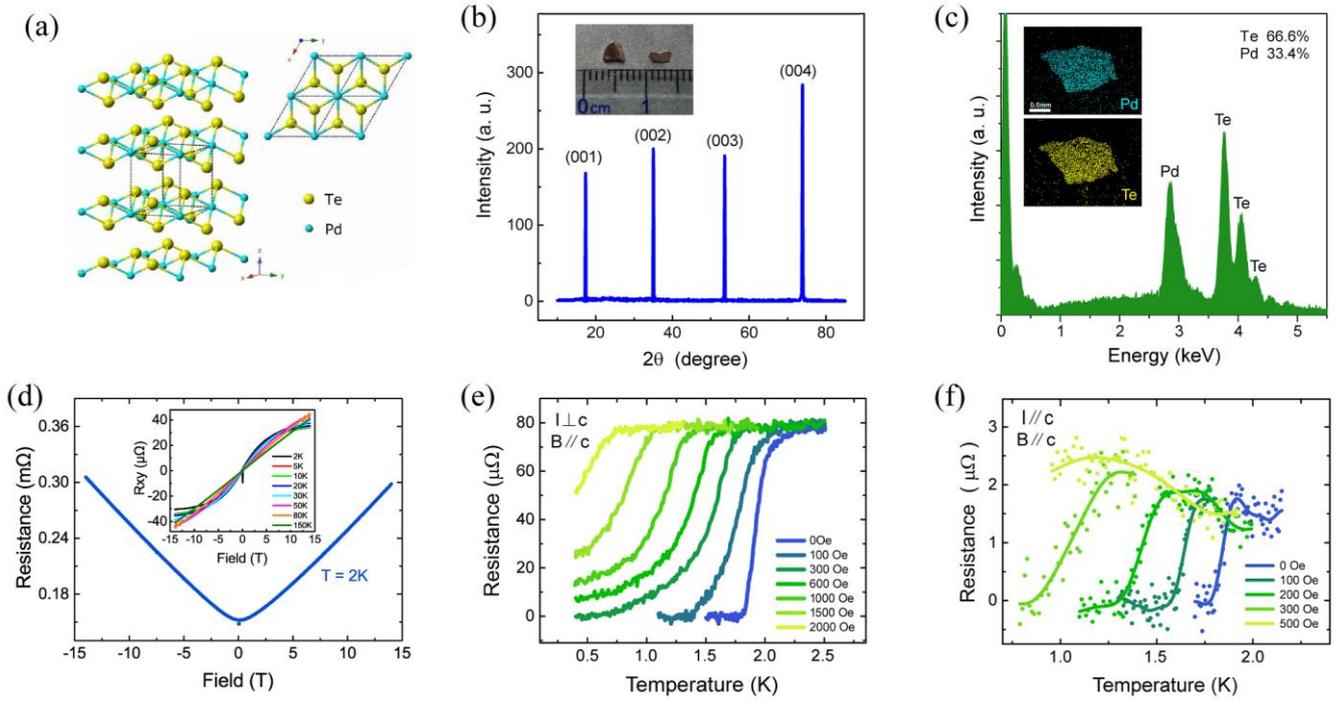

FIG. 1. PdTe$_2$ Crystal growth and electrical transport under magnetic fields.

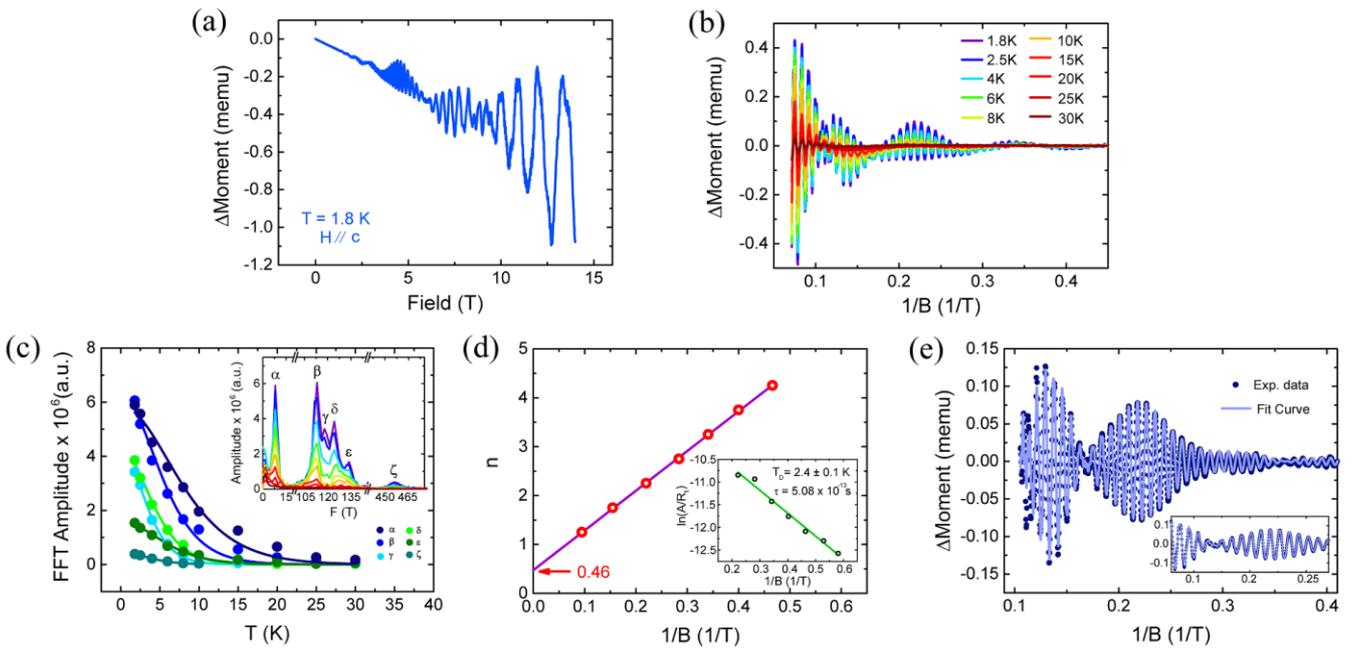

FIG. 2. The dHvA oscillations and nontrivial Berry phase.

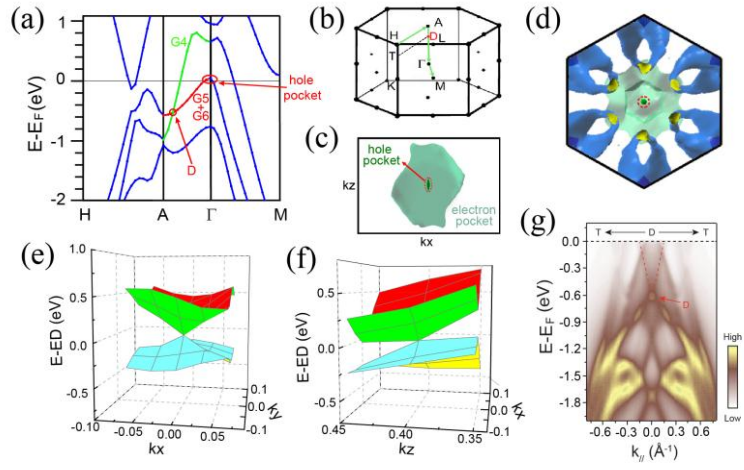

FIG. 3. Matching the dHvA components in the calculations of type-II Dirac cone.

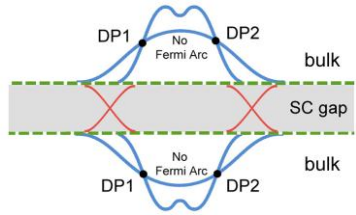

FIG. 4. Optimizing the topological superconductor carrier (TSC) by three-dimensional semimetals.

TABLES

TABLE I. Parameters extracted from the dHvA oscillations ($H \parallel c$) of $PdTe_2$. $F$, oscillation frequency; $S_f$, crosssection of Fermi surface; $k_f$, Fermi wave vector; $T_D$, Dingle temperature; $m^*/m_0$, relative effective mass; $\tau$, quantum relaxation time; $\mu_q$, quantum mobility.

| $F$ (T) | $S_f$ (Å$^{-2}$) | $k_f$ (Å$^{-1}$) | $T_D$ (K) | $m^*/m_0$ | $\tau$ (ps) | $\mu_q$ (cm$^2$/Vs) |
|---|---|---|---|---|---|---|
| 8.0 | $7.67 \times 10^{-4}$ | $1.56 \times 10^{-2}$ | 2.4 | 0.14 | 0.51 | 6209 |
| 113.2 | $1.09 \times 10^{-2}$ | $5.88 \times 10^{-2}$ | 3.4 | 0.21 | 0.36 | 3030 |
| 117.9 | $1.13 \times 10^{-2}$ | $6.00 \times 10^{-2}$ | 1.9 | 0.33 | 0.65 | 3464 |
| 124.3 | $1.19 \times 10^{-2}$ | $6.16 \times 10^{-2}$ | 3.0 | 0.26 | 0.41 | 2722 |
| 133.9 | $1.28 \times 10^{-2}$ | $6.39 \times 10^{-2}$ | 6.8 | 0.20 | 0.18 | 1593 |
| 455.8 | $4.34 \times 10^{-2}$ | 0.118 | 5.7 | 0.29 | 0.21 | 1293 |